\documentclass{rspublic}
\usepackage{amssymb,amsmath}
\usepackage{graphicx}

\begin{document}

\title[Landau-Lifshitz-Gilbert Equation]{The Fascinating World of Landau-Lifshitz-Gilbert Equation: An Overview}

\author[Lakshmanan]{M. Lakshmanan}

\affiliation{Centre for Nonlinear Dynamics, Department of Physics, \\Bharathidasan Univeristy, Tiruchirapalli - 620 024, India}

\label{firstpage}

\maketitle

\begin{abstract}{LLG Equation, Spin systems, Integrability, Chaos and Patterns}

The Landau-Lifshitz-Gilbert (LLG) equation is a fascinating nonlinear evolution equation both from mathematical  and physical
points of view.  It is related to the dynamics of several important physical systems such as ferromagnets, vortex filaments, 
moving space curves, etc. and has intimate connections with many of the well known integrable soliton equations, including 
nonlinear Schr\"odinger and sine-Gordon equations.  It can admit very many dynamical structures including spin waves, elliptic
function waves, solitons, dromions, vortices, spatio-temporal patterns, chaos, etc. depending on the physical and spin dimensions
and the nature of interactions.  An exciting recent development is that the spin torque effect in nanoferromagnets is described
by a generalization of the LLG equation which forms a basic dynamical equation in the field of spintronics.  This article will briefly 
review these developments as a tribute to Robin Bullough who was a great admirer of the LLG equation.

\end{abstract}
\section{Introduction}
Spin systems generally refer to ordered magnetic systems.  Specifically, spin angular momentum or spin is   an intrinsic property associated with quantum particles, which does not have a classical counterpart.  Macroscopically, all substances are magnetic to some extent and every material when placed in a magnetic field acquires a magnetic moment or magnetization.  In analogy with the relation between the dipole moment of a current loop in a magnetic field and orbital angular momentum of a moving electron, one can relate the magnetic moment/magnetization with the expectation value of the spin angular momentum operator, which one may call simply as spin.  In ferromagnetic materials, the moment of each atom and even the average is not zero.  These materials are normally made up of domains, which exhibit long range ordering that causes the spins of the atomic ions to line up parallel to each other in a domain.  The underlying interaction  (Hillebrands \& Ounadjela 2002) originates from a spin-spin exchange interaction that is caused by the overlapping of electronic wave functions.  Additional interactions which can influence the magnetic structures include magnetocrystalline anisotropy, applied magnetic field, demagnetization field, biquadratic exchange and other weak interactions.  Based on phenomenological grounds, by including effectively the above type of interactions, Landau and Lifshitz (1935) introduced the basic dynamical equation for magnetization or spin $\vec S(\vec r,t)$ in bulk materials, where the effect of relativistic interactions were also included as a damping term.  In 1954, Gilbert (2004) introduced a more convincing form for the damping term, based on a Lagrangian approach, and the combined form is now called the Landau-Lifshitz-Gilbert (LLG) equation, which is a fundamental dynamical system in applied magnetism (Hillebrands \& Ounadjela 2002; Mattis 1988; Stiles \& Miltat 2006).

The LLG equation for the unit spin vector $\vec S(\vec r,t)$, because of the constancy of length, is a highly nonlinear partial differential equation in its original form for bulk materials.  Depending on the nature of the spatial dimensions and interactions, it can exhibit a very large variety of nonlinear structures such as spin waves, elliptic function waves, solitary waves, solitons, lumps, dromions, bifurcations and chaos, spatiotermporal patterns, etc.  It exhibits very interesting differential geometric properties and has close connections with many integrable soliton and other systems, for special types of interactions.  In the general situations, the system is highly complex and nonintegrable.  Both from physical and mathematical  points of view its analysis is highly challenging but rewarding.

One can also deduce the LLG equation starting from a lattice spin Hamiltonian, by postulating appropriate Poisson brackets, and writing down the corresponding evolution equations and then introducing the Gilbert damping term phenomenologically.  Thus we can have LLG equation for a single spin, a lattice of spins and then the continuum limit in the form of nonlinear ordinary differential equation(ODE), a system of coupled nonlinear ODEs and a nonlinear partial differential equation, respectively, for the unit spin vector(s).  Analysis of the LLG equation for discrete spin systems turns out to be even harder than the continuum limit due to the nature of nonlinearity.  However, apart from exact analytic structures, one can also realize the onset of bifurcations, chaos and patterns more easily in discrete cases.  Thus the LLG equation turns out to be an all encompassing nonlinear dynamical system.

The LLG equation has also close relationship with several other physical systems, for example motion of a vortex filament, motion of curves and surfaces, $\sigma $-models in particle physics, etc.  One of the most exciting recent developments is that a simple generalization of the LLG equation also forms the basis of the so called spin torque effect in nanoferromagnets in the field of spintronics.

With the above developments in mind, in this article we try to present a brief overview of the different aspects of the LLG equation, concentrating on its nonlinear dynamics.  Obviously the range of LLG equation is too large and it is too difficult to cover all aspects of it in a brief article and so the presentation will be more subjective.  The structure of the paper will be as follows.  In Sec.2, starting from the dynamics of the single spin, extension is made to a lattice of spins and continuum systems to obtain the LLG equation in all the cases.  In Sec.3, we briefly point out the spin torque effect.  Sec.4, deals with exact solutions of discrete spin systems, while Sec.5 deals with continuum spin systems in (1+1) dimensions and magnetic soliton solutions.  Sec.5 deals with (2+1) dimensional continuum spin systems.  Concluding remarks are made in Sec.6.                                                       
\section{Macroscopic Dynamics of Spin Systems and LLG Equation}
\label{sec1}
To start with, in this section we will present a brief account of the phenomenological derivation of the Landau-Lifshitz-Gilbert (LLG) equation starting with the equation of motion of a magnetization vector in the presence of an applied magnetic field (Hillebrands \& Ounadjela, 2002).  Then this analysis is extended to the case of a lattice of spins and its continuum limit, including the addition of a phenomenological damping term, to obtain the LLG equation.

\subsection{Single Spin Dynamics}
\label{sec2}
Consider the dynamics of the spin angular momentum operator $ \vec S $ of a free electron under the action of a time-dependent external magnetic 
field with the Zeeman term given by the Hamiltonian
\begin{align}
H_s = - \frac{g \mu_{B}}{\hbar}~ \vec{S} . \vec{B} (t),~~~~~\vec{B} (t) = \mu_{0} \vec{H} (t), \tag{1}
\label{met1}
\end{align}
where g, $ \mu_{B}$ and $\mu_{0}$ are the gyromagnetic ratio, Bohr magneton and permeability in vacuum, respectively.  Then, from the Schr\"odinger equation, the expectation value of the spin operator can be easily shown to satisfy the dynamical equation, using the angular momentum commutation relations, as
\begin{align}
\frac{d}{dt} <\vec {S}(t)> ~=~ \frac{g \mu_{B}}{\hbar}~ < \vec {S}(t)>  \times \vec{B}(t). \tag{2}
\end{align}
Now let us consider the relation between the classical angular momentum $\vec L $ of a moving electron and the dipole moment $\mathcal{\vec M}_{e} $ of a current loop immersed in  a uniform magnetic field, $ \mathcal{\vec M}_{e} = \frac{e}{2m} ~ \vec L$, where $e$ is the charge and $m$ is the mass of the electron.  Analogously one can define the magnetization $\mathcal{\vec M} = \frac{g \mu_{B}}{\hbar}<\vec{S}> \equiv \gamma <\vec S> $, where $\gamma=\frac{g \mu_{B}}{\hbar}$.  Then considering the magnetization per unit volume, $\vec M$, from (2) one can write the evolution equation for the magnetization as
\begin{align}
\frac {d\vec M}{dt}~=~ -\gamma_{0}[\vec {M}(t) \times \vec{H} (t)],\tag{3}
\end{align}
where $\vec B~=~ \mu_0 \vec H $ and $\gamma_0~=~\mu_{0} \gamma.$

From (3) it is obvious that $\vec M . \vec M$~=~constant and $\vec M . \vec H$ ~=~ constant. Consequently the magnitude of the magnetization vector remains constant in time, while it precesses around the magnetic field $\vec H$ making a constant angle with it.  Defining the unit magnetization vector
\begin{align}
\vec{S} (t)~=~ \frac{\vec{M}(t)}{|\vec{M}(t)|}, ~~\vec {S}^2 = 1,~~\vec S = (S^x,S^y,S^z), \tag{4}
\end{align}
which we will call simply as spin hereafter, one can write down the spin equation of motion as (Hillebrands \& Ounadjela 2002)
\begin{align}
\frac{d\vec{S}(t)}{dt}~=~-\gamma_{0}[\vec {S}(t) \times \vec {H}(t)],~~ \vec H = (H^x,H^y,H^z) \tag{5}
\end{align}
and the evolution of the spin can be schematically represented as in Fig. 1(a).
\begin{figure}
\centering\includegraphics[width=0.5 \linewidth]{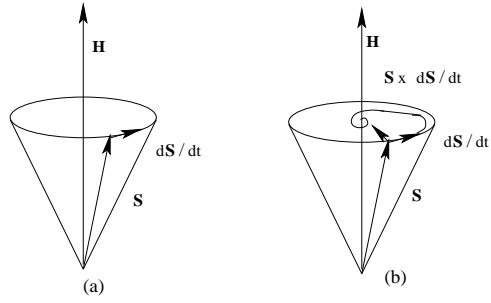}
\caption{Evolution of a single spin (a) in the presence of a magnetic field and (b) when damping is included.}
\end{figure}

It is well known that experimental hysteresis curves of ferromagnetic substances clearly show that beyond certain critical values of the applied magnetic field, the magnetization saturates, becomes uniform and aligns parallel to the magnetic field.  In order to incorporate this experimental fact, from phenomenological grounds one can add a damping term suggested by Gilbert (2004) so that the equation of motion can be written as

\begin{align}
\frac{d\vec{S}(t)}{dt} &=~-\gamma_{0}[\vec {S}(t) \times \vec {H}(t)]+\lambda \gamma_{0}\left[\vec S \times \frac {d\vec S}{dt}\right], \lambda \ll 1 (damping~ parameter). \tag{6} 
\end{align}
On substituting the expression (6) again for $\frac{d\vec S}{dt}$ in the third term of (6), it can be rewritten as 
\begin{align}
(1+\lambda^2 \gamma_{0})\frac{d\vec S}{dt}~=~-\gamma_{0} [\vec S \times \vec H (t)] - \lambda \gamma_{0} ~ \vec S \times [\vec S \times \vec H(t)]. \nonumber
\end{align}
After a suitable rescaling of $t$, Eq.(6) can be rewritten as
\begin{align}
\frac{d\vec S}{dt} & =~(\vec{S} \times \vec {H}(t))+\lambda \vec{S} \times [\vec {S} \times \vec {H}(t)] \nonumber \\
& =~\vec{S} \times \vec {H}_{eff}. \tag{7}
\end{align}
Here the effective field including damping is
\begin{align}
\vec {H}_{eff}~=~\vec{H}(t) + \lambda [\vec {S} \times \vec {H}]. \tag{8}
\end{align}
The effect of damping is shown in Fig 1(b).  Note that in Eq.(7) again the constancy of the length of the spin is maintained.  Eq.(7) is the simplest form of Landau-Lifshitz-Gilbert equation, which represents the dynamics of a single spin in the presence of an applied magnetic field $\vec{H}(t)$. 

\subsection{Dynamics of Lattice of Spins and Continuum Case}
\label{sec3}
 The above phenomenological analysis can be easily extended to a lattice of spins representing a ferromagnetic material.  For simplicity, considering an one dimensional lattice of N spins with nearest neighbour interactions, onsite anisotropy, demagnetizing field, applied magnetic field, etc., the dynamics of the $ith$ spin can be written down in analogy with the single spin as the LLG equation,
\begin{align}
\frac{d\vec{S}_i}{dt}~=~\vec {S}_i \times \vec {H}_{eff},~~~ i = 1,2,......N \tag{9}
\end{align}
where
\begin{align}
\vec{H}_{eff}=~&(\vec{S}_{i+1} + \vec{S}_{i-1} + A S^x_i \vec n_x + B S^y_i \vec {n}_y + C S^z_i \vec n_z + \vec H(t) +......) \nonumber \\
 &-\lambda \{\vec{S}_{i+1} + \vec{S}_{i-1} + A S^x_i \vec n_x + B S^y_i \vec {n}_y + C S^z_i \vec n_z + \vec H(t) +......\} \times \vec S_i \tag{10}
\end{align} 
Here $A,B,C$ are anisotropy parameters, and $\vec n_x, \vec n_y, \vec n_z $ are unit vectors along the $x,y$ and $z$ directions, respectively.  One can include other types of interactions like biquadratic exchange, spin phonon coupling, dipole interactions, etc. Also Eq.(9) can be generalized to the case of square and cubic lattices as well, where the index $i$ has to be replaced by the appropriate lattice vector $\vec i$.

In the long wavelength and low temperature limit, that is in the continuum limit, one can write 
\begin{align}
\vec S_{\vec i} (t) ~&=~ \vec S(\vec r, t),~~~~~\vec r = (x,y,z), \nonumber \\
\vec S_{\vec i + \vec 1} + \vec S_{\vec i - \vec 1}~&=~ \vec S(\vec r, t) + \vec a. \vec \nabla \vec S + \frac{a^2}{2} \nabla ^2 \vec S + higher~orders, \tag{11}
\end{align}
$(\vec a$: lattice vector) so that the LLG equation takes the form of a vector nonlinear partial differential equation (as $\vec a \rightarrow \vec {0}),$
\begin{align}
\frac{\partial \vec S(\vec r,t)}{\partial t} = &\vec S \times \left[\left\{\nabla^2 \vec S+AS^x \vec n_x+BS^y \vec n_y+CS^z \vec n_z+ \vec H(t)+....\right\}\right] \nonumber \\
&+\lambda \left[\left\{\nabla^2 \vec S+AS^x \vec n_x+BS^y \vec n_y+CS^z \vec n_z+ \vec H(t)+....\right\}\times \vec S(\vec r,t)\right],\nonumber \\
= &\vec S \times \vec H_ {eff}  \tag{12}\\
\vec S(\vec r,t)=&(S^x(\vec r,t),S^y(\vec r,t),S^z(\vec r,t)),~~~ \vec S^2 = 1. \tag{13}
\end{align}
In fact, Eq.(12) was deduced from phenomenological grounds for bulk magnetic materials by Landau and Lifshitz originally in 1935.
\subsection{Hamiltonian Structure of the LLG Equation in the Absence of Damping}
\label{sec4}
The dynamical equations for the lattice of spins (9) in one dimension (as well as in higher dimensions) posses a Hamiltonian structure in the absence of damping.

Defining the spin Hamiltonian
\begin{align}
H_s = -\sum_{\{i,j\}} \vec S_i . \vec S_ {i+1} + A (S^x_i)^2 + B(S^y_i)^2 + C(S^z_i)^2 + \mu (\vec H(t).\vec S_i) + ..... \tag{14}
\end{align}
and the spin Poisson brackets between any two functions of spin $\mathcal{A}$ and $\mathcal{B}$ as
\begin{align}
\{ \mathcal{A},\mathcal{B} \} =  \sum_{\alpha,\beta,\gamma=1}^3  \in_{\alpha \beta \gamma} \frac{\partial \mathcal{A}}{\partial S^\alpha} \frac{\partial \mathcal{B}}{\partial S^\beta} S^\gamma ,\tag{15}
\end{align}
one can obtain the evolution equation (9) for $\lambda=0$ from $\frac{d\vec{S}_i}{dt}= \{\vec{S}_i,H_s\}$.  Here $ \in_{\alpha\beta\gamma} $ is the Levi-Civitta tensor. Similarly for the continuum case, one can define the spin Hamiltonian density
\begin{align}
H_s = \frac{1}{2} [ (\nabla \vec S)^2 + A(S^x)^2 + B (S^y)^2 +C(S^z)^2 + (\vec H . \vec S) + H_{demag}+......] \tag{16}
\end{align}
along with the Poisson  bracket relation
\begin{align}
\left\{S^\alpha (\vec r,t), S^\beta(\vec r^{'},t^{'})\right\}= \in_{\alpha\beta\gamma} S^\gamma \delta(\vec r-\vec r^{'}, t-t^{'}),  \tag{17}
\end{align}
and deduce the spin field evolution equation (12) for $\lambda=0$.

Defining the energy as
\begin{align}
E = \frac{1}{2} \int d^3 r[(\nabla \vec S)^2 + A (S^x)^2 +B(S^y)^2 +C(S^z)^2+ \vec H .\vec S+....]  \tag{18}
\end{align}
one can easily check that 
\begin{align}
\frac{dE}{dt}  =-\lambda \int  |\vec S_t|^2 d^3 r.\nonumber
\end{align}
Then when $\lambda > 0$, the system  is dissipative, while for $\lambda =0$ the system is conservative.
\section{Spin Torque Effect and the Generalized LLG Equation}
\label{}
Consider the dynamics of spin in a nanoferromagnetic film under the action of a spin current (Stiles \& Miltat, 2006; Bertotti $et~al$. 2009).  In recent times it has been realized that if the current is spin polarized, the transfer of a strong current across the film results in a transfer of spin angular momentum to the atoms of the film.  This is called spin torque effect and forms one of the basic ideas of the emerging field of spintronics.  The typical set up of the nanospin valve pillar consists of two ferromagnetic layers, one a long ferromagnetic pinned layer, and the second one is of a much smaller length, separated by a spacer conductor layer [6], all of which are nanosized.  The pinned layer acts as a reservoir of spin polarized current which on passing through the conductor and on the ferromagnetic layer induces an effective torque on the spin magnetization in the ferromagnetic film, leading to rapid switching of the spin direction of the film.  Interestingly, from a semiclassical point of view, the spin transfer torque effect is described by a generalized version of the LLG equation (12), as shown by Berger (1996), and by Slonczewski (1996), in 1996.  Its form reads
\begin{align}
\frac{\partial \vec S}{\partial t} = \vec S \times [\vec H_{eff} + \vec S \times \vec j],~~~\vec S=(S^x, S^y, S^z),~~~ \vec S^2=1 \tag{19}
\end{align}
where the spin current  term
\begin{align}
\vec j = \frac{a.\vec S_P}{f(P)(3+\vec S . \vec S_P)},~~~~~f(P)=\frac{(1+P^3)}{4P^{\frac{3}{2}}}. \tag{20}
\end{align}
Here $ \vec S_P$ is the pinned direction of the polarized spin current which is normally taken as perpendicular to the direction of flow of current, $a$ is a constant factor related to the strength of the spin current and $f(P)$ is the polarization factor deduced by Slonczewski (1996) from semiclassical arguments.  From an experimental point of view  valid for many ferromagneic materials it is argued that  it is sufficient to approximate the spin current term simply as
\begin{align}
\vec j~=~a \vec S_p  \tag{21}
\end{align}
so that LLG equation for the spin torque effect can be effectively written down as
\begin{align}
\frac{\partial \vec S}{\partial t} = \vec S \times [\vec H_{eff} + a \vec S\times \vec S_p],  \tag{22}
\end{align}
where
\begin{align}
\vec H_{eff}=(\nabla ^2 \vec S + A S^x \vec i + B S^y \vec j + C S^z \vec k + \vec H_{demag} + \vec H(t) + .....) - \lambda \vec S \times \frac {\partial \vec S}{\partial t}. \tag{23}
\end{align}
Note that in the present case, using the energy expression (18), one can prove that
\begin{align}
\frac{dE}{dt} = \int [-\lambda |S_t|^2 + a(\vec S_t \times \vec S).\vec S_p] d^3 r. \tag{24}
\end{align}
This implies that energy is not necessarily decreasing along trajectories.  Consequently, many interesting dynamical features of spin can be expected to arise in the presence of the spin current term.

In order to realize these effects more transparently, let us rewrite the generalized LLG equation(22) in terms of the complex stereographic variable $\omega(\vec r,t)$(Lakshmanan \& Nakumara, 1984) as
\begin{align}
\omega ~=~ &\frac{S^x + i S^y}{(1+S^z)}, \tag{25a}\\ 
S^x = \frac{\omega+\omega^*}{(1+\omega\omega^*)},~~~~~S^y =&~ \frac{1}{i} \frac{(\omega - \omega^*)}{(1+\omega \omega^*)},~~~~~S^z = \frac{(1- \omega \omega^*)}{(1+\omega \omega^*)} \tag{25b}
\end{align}
so that Eq.(20) can be rewritten (for simplicity $\vec H_{demag}=0$ )
\begin{align}
i(1-i\lambda)\omega_t &+ \nabla^2 \omega - \frac{2 \omega^* (\nabla \omega)^2}{(1+\omega \omega^*)} + \frac{A}{2} \frac{(1-\omega^2)(\omega+\omega^*)}{(1+\omega\omega^*)}\nonumber\\ &+ \frac{B}{2} \frac {(1+\omega^2)(\omega-\omega^*)}{(1+\omega\omega^*)} - C(\frac{1-\omega\omega^*}{1+\omega\omega^*})\omega\nonumber \\ +\frac{1}{2}(H^x - i j^x)(1-\omega^2) &+ \frac{1}{2} i (H^y+ij^y) (1+\omega^2) - (H^z + ij^z)\omega = 0,  \tag{26}
\end{align}
where $\vec j= a \vec S_p,~ \omega_t = (\frac{\partial \omega}{\partial t}).$

It is clear from Eq.(26) that the effect of the spin current term $\vec j$ is simply to change the magnetic field $\vec H = (H^x,H^y,H^z)$ as $(H^x-ij^x, H^y+ij^y, H^z+ij^z)$. Consequently the effect of the spin current is effectively equivalent to a magnetic field, though complex.  The consequence is that the spin current can do the function of the magnetic field perhaps in a more efficient way because of the imaginary term.

To see this in a simple situation, let us consider the case of a homogeneous ferromagnetic film so that there is no spatial variation and the anisotropy and demagnetizing fields are absent, that is we have (Murugesh \& Lakshmanan, 2009a)
\begin{align}
(1-i\lambda)\omega_t = -(a-iH^z) \omega.  \tag{27}
\end{align}
Then on integration one gets
\begin{align}
\omega(t) &= \omega(0) exp[\frac{-(a-iH^z)t}{(1-i\lambda)}] \nonumber\\
&= \omega(0) exp\left[-\frac{(a+\lambda H^z)t}{(1+|\lambda|^2)}\right] exp\left[-i\frac{(a\lambda -H^z)t}{(1+|\lambda|^2)}\right].  \tag{28}
\end{align}
Obviously the first exponent describes a relaxation or switching of the spin, while the second term describes a precession.  From the first exponent in (28), it is clear that the time scale of switching is given by $ \frac {1+|\lambda|^2}{(a+\lambda H^z)}$.  Here $\lambda$ is small which implies that the spin torque term is more effective in switching the magnetization vector.  Furthermore letting $H^z$ term to become zero, we note that in the presence of damping term the spin transfer produces the dual effect of precession and dissipation.

In Fig.2, we point out clearly how the effect of spin current increases the rate of switching of the spin even in the presence of anisotropy.  Further, one can show that interesting bifurcation scenerio, including period doubling bifurcations to chaotic behaviour, occurs on using a periodically varying applied magnetic field in the presence of a constant magnetic field and constant spin current (Murugesh \& Lakshmanan, 2009a). Though a periodically varying spin current can also lead to such a bifurcations-chaos scenerio (Yang $et~al$. 2007), we believe the technique of
applying a periodic magnetic field in the presence of constant spin current is much more feasible experimentally.  To realize this one can take (Murugesh \& Lakshmanan, 2009b)
\begin{align}
\vec H_{eff} = \kappa(\vec S.\vec e_{\|}) \vec e_{\|} + \vec H_{demag} + \vec H(t),  \tag{29}
\end{align}
where $\kappa$ is the anisotropy parameter and $\vec e_{\|}$ is the unit vector along the anisotropy axis and
\begin{align}
\vec H_{demag} = -4 \pi (N_1 S^x \vec i + N_2 S^y \vec j + N_3 S^z \vec k).  \tag{30}
\end{align}
Choosing
\begin{align}
\vec H(t) = (0,0,H^z), ~~~~\vec e_{\|} = (sin\theta_{\|} cos\phi_{\|}, sin\theta_{\|} sin\phi_{\|}, cos\theta_{\|}) , \tag{31}
\end{align}
the LLG equation in stereographic variable can be written down(in the absence of exchange term) as (Murugesh \& Lakshmanan, 2009a,b)
\begin{align}
(1-i\lambda)\omega_t = &-\gamma(a-i H^z)\omega + iS_{\|}\kappa\gamma\left[cos \theta_{\|}\omega - \frac{1}{2} sin\theta_{\|}\left(e^{i\phi_{\|}}-\omega^2 e^{-i\phi_{\|}}\right)\right]  \nonumber\\
&- \frac{i4\pi\gamma}{(1+|\omega|^2)}\left[N_3 (1-|\omega|^2)\omega - \frac{N_1}{2}(1-\omega^2-|\omega|^2)\omega  \right.\nonumber\\
&\left. - \frac{N_2}{2} (1+\omega^2-|\omega|^2)\omega-\left(\frac{N_1-N_2}{2}\right) \bar{\omega} \right], \tag{32}
\end{align}
where $S_{\|} = \vec S. \vec e_{\|}$.  Solving the above equation numerically one can shown that Eq.(27) exhibits a typical period doubling bifurcation route to chaos as shown in Fig.3.

\begin{figure}
\centering\includegraphics[width=0.5 \linewidth]{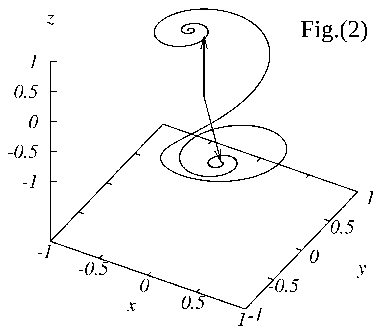}\includegraphics[width=0.5 \linewidth]{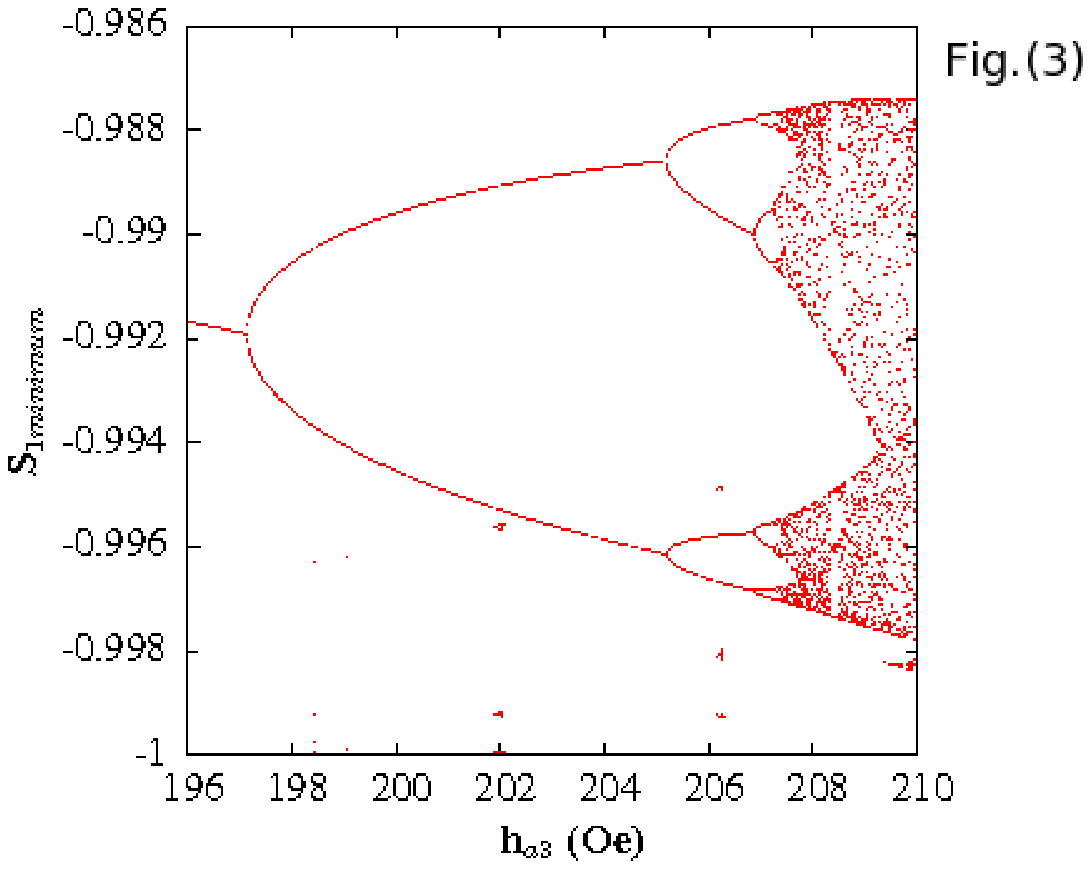} 
Fig.2:Switching of spin due to spin current in the presence of anisotropy (Murugesh \& Lakshmanan, 2009a).\\ Fig.3: Bifurcation diagram corresponding to Eq.(32). Here $h_{{a_3}}=H^z$ (Murugesh \& Lakshmanan, 2009b).
\end{figure}

The existence of periodic and chaotic spin oscillations in a homogeneous nano-spin transfer oscillator (STO) leads to exciting other possibilities.  For example, an array/network of STO's can lead to the possibility of synchronized microwave power or synchronized chaotic oscillations (Grollier $et~al$. 2006).  Such studies are in progress.  Other possibilities include the study of inhomogeneous films (including spatial variations) and discrete lattices, including higher dimensions (Bazaliy $et~al$. 2004).

\section{Anisotropic Heisenberg Spin Lattice}
\label{}
Next we consider the dynamics of a discrete anisotropic Heisenberg spin systems without damping.  
Consider the Hamiltonian
\begin{align}
H= -\sum_n (A S^x_n S^x_{n+1}+ B S^y_n S^y_{n+1} + C S^z_n S^z_{n+1}) - D\sum(S^z_n)^2 - \vec H . \sum \vec S_n \tag{33}
\end{align}
so that the equation of motion becomes(using the Poisson bracket relations(15))
\begin{align}
\frac{d\vec S_n}{dt} = \vec S_n \times &[A(S^x_{n+1} + S^x_{n-1})\vec i+B(S^y_{n+1} +S^y_{n-1})\vec j \nonumber \\
&+ C(S^z_{n+1}+S^z_{n-1})\vec k + 2D S^z_n \vec k] + \vec S_n \times \vec H, \tag{34}\\
& n = 1,2,3,...N. \nonumber
\end{align}
Recently (Lakshmanan \& Saxena, 2008; Roberts \& Thompson, 1988) it has been found that the coupled system (34) admits several classes of exact solutions, though the system may not be completely integrable for any choice of the parameters, including the pure isotropic one $(A=B=C=1, D=0, \vec H=0).$

The class of exact solutions to (34) are as follows.

\subsection{Spatially homogeneous time dependent solution:}

\begin{align}
S^x_{n}  &= -\sqrt{1-\gamma^2 k^{'2}} ~~~ sn(\omega {t} + \delta , k), \tag{35a} \\
S^y_n &= \sqrt{1-\gamma^2} ~~~cn(\omega {t} + \delta, k),\tag{35b}\\
S^z_n &= \gamma ~~~dn(\omega {t}+\delta, k),\tag{35c}
\end{align}
where the frequency
\begin{align}
\omega = 2\gamma&\sqrt{(B-C)(A-C)},~~~~~k^2 = \frac{1-\gamma^2}{\gamma^2} \frac {(B-A)}{(A-C)} ~~~(B>A>C).  \tag{35d}
\end{align}
Here $\gamma$ and $\delta$ are arbitrary parameter and $k$ is the modulus of the Jacobian elliptic functions.
\subsection{Spatially oscillatory time periodic solutions}
\label{}
\begin{align}
S^x_n &= (-1)^{n+1} \sqrt{1-\gamma^2 k^{'2}}~sn(\omega t + \delta ,k), \tag{36a} \\
S^y_n &= (-1)^n \sqrt{1-\gamma^2}~cn(\omega t + \delta ,k), \tag{36b} \\
S^z_n &= \gamma~dn(\omega t + \delta ,k), \tag{36c} 
\end{align}
where
\begin{align}
\omega = 2\gamma\sqrt{(A+C)(B+C)},~~~~~k^2 = \frac{1-\gamma^2}{\gamma^2} \frac {(B-A)}{(A+C)}. \tag{36d}
\end{align}
This solution corresponds to a nonlinear magnon.
\subsection{Linear magnon solutions}
In the uniaxial anisotropic case $A = B < C$, the linear magnon solution is
\begin{align}
S^x_n &= \sqrt{1-\gamma^2 }~sin(pn-\omega t + \delta ), \tag{37a} \\
S^y_n &= \sqrt{1-\gamma^2}~cos(pn-\omega t + \delta ),  \tag{37b} \\
S^z_n &= \gamma \tag{37c} 
\end{align}
with the dispersion relation
\begin{align}
\omega = 2\gamma(C -A~ cos~ p). \tag{37d}
\end{align}
\subsection{Nonplanar static structures for XYZ and XYY models}
\label{}
In this case we have the periodic structures
\begin{align}
S^x_n &= \sqrt{1-\gamma^2 k^{'2}}~sn(pn+ \delta ,k), \tag{38a} \\
S^y_n &= \sqrt{1-\gamma^2}~cn(pn+ \delta ,k ),  \tag{38b} \\
S^z_n &= \gamma~dn(pn+ \delta ,k ),\tag{38c} 
\end{align}
where
\begin{align}
k^2 = \frac{A^2 - B^2}{A^2 - C^2},~~~~~dn(p,k) = \frac{B}{A}. \tag{38d}
\end{align}
In the limiting case $k=1$, one can obtain the localized single soliton (solitary wave) solution
\begin{align}
S^x_n = tanh(pn+\delta),~~~S^y_n = \sqrt{1-\gamma^2}~ sech(pn+\delta),~~~S^z_n = \gamma~sech(pn+\delta).  \tag{39}
\end{align}
\subsection{Planar (XY) Case:}
\begin{align}
Case(i):  ~~~S^x_n = sn(pn+\delta,k),~~~S^y_n = cn(pn+\delta,k),~~~S^z_n = 0,  \tag{40}
\end{align}
where $dn(p,k) = \frac{B}{A}.$ In the limiting case, $k=1$, 
we have the solitary wave solution
\begin{align}
S^x_n= tanh(pn+\delta,k),~~~S^y_n = sech(pn+\delta),~~~S^z_n = 0.  \tag{41}
\end{align}
\begin{align}
Case(ii):~~~S^x_n = k~sn(pn+\delta,k),~~~S^z_n = dn(pn+\delta,k),~~~S^y_n = 0, \tag{42}
\end{align}
where $cn(p,k)=\frac{C}{A}$.  In the limit $k=1$, we have
\begin{align}
S^x_n = tanh(pn+\delta),~~~S^y_n = 0,~~~S^z_n = sech(pn+\delta). \tag{43}
\end{align}
\subsection{Nonplanar XYY Structures:}
We have
\begin{align}
S^x_n = cn(pn+\delta,k),~~~S^y_n = \gamma~sn(pn+\delta,k),~~~
S^z_n = \sqrt{1-\gamma^2}~sn(pn+\delta,k), \tag{44}
\end{align}
where $dn(p,k) = \frac{A}{B}.$  In the limiting case, we have the domain wall structure
\begin{align}
S^x_n = sech(pn+\delta),~~~S^y_n = \gamma~tanh(pn+\delta),~~~S^z_n = \sqrt{1-\gamma^2}~tanh(pn+\delta).  \tag{45}
\end{align}
In all the above cases one can evaluate the energies associated with the different structures and their linear stability properties.  For details, one may refer to (Lakshmanan \& Saxena, 2008).
\subsection{Solutions in the presence of onsite anisotropy and constant external magnetic field}
\subsubsection{Onsite anisotropy, $D \neq 0,~H=0,~A,B,C \neq 0:$}
All the three types of solutions (35), (36) and (37) exist here also, except that the parameter $C$ has to be replaced by $(C-D)$ in each of these equations on their right hand sides.

\subsubsection{Constant external field case, $\vec H = (H_x,0,0),~ B=C \neq A,~ D=0.$}
An exact solution is 
\begin{align}
S^x_n &= sn(pn+\delta,k), \tag{46a}\\
S^y_n &= sin(\omega t+\gamma)~cn(pn+\delta,k),  \tag{46b}\\
S^z_n &= cos(\omega t+\gamma)~cn(pn+\delta,k), \tag{46c}
\end{align}
where $dn(p,k) = \frac{C}{A},$ and $\omega = H_x.$

One can study the linear stability of static solutions and investigate the existence of the so called Peierls-Nabarro barrier, that is whether the total lattice energy depends on the location of the soliton or not, for details see (Lakshmanan \& Saxena, 2008).
\subsection{Integrability of the static case }
Granovskii and Zhedanov(1986) have shown that the static case of the pure anisotropic system
\begin{align}
\vec S_n \times \left[A(S^x_{n+1} + S^x_{n-1}) +B(S^y_{n+1} + S^y_{n-1})+C(S^z_{n+1} + S^z_{n-1})\right] = 0 \tag{47}
\end{align}
is equivalent to a discretized version of the Schr\"odinger equation with two level Bargmann type potential or a discrete analog of Neumann system  (Veselov, 1987) and is integrable.
\subsection{Ishimori spin chain:}
There exists a mathematically interesting spin chain which is completely integrable and which was introduced by Ishimori (1982).  Starting with a Hamiltonian

$H = -log(1+ \sum {\vec S_n . S_{n+1}})$, the spin equation becomes
\begin{align}
\dot S_n = \vec S_n \times \left[ \frac{\vec S_{n+1}}{1+ \vec S_n .\vec S_{n+1}} + \frac {\vec S_{n-1}}{1+ \vec S_n . \vec S_{n-1}}\right] \tag{48}
\end{align}
It admits a Lax pair and so is completely integrable. However, no other realistic spin system is known to be completely integrable.  It is interesting to note that Eq.(48) also leads to an integrable reversible map.  Assuming a simple time dependence,
\begin{align}
\vec S_n(t) = (cos \phi_n cos \omega t, cos \phi_n sin \omega t, sin \phi_n),  \tag{49}
\end{align}
Quispel, Robert and Thompson (1988) have shown that Eq.(48) reduces to the integrable map
\begin{align}
x_{n+1} = &[2x^3_n + \omega x^2_n + 2 x_n - \omega - x_{n-1}(-x^4_n - \omega x^3_n + \omega x_n +1)] \nonumber \\
&\times [-x^4_n - \omega x^3_n + \omega x_n +1 - x_{n-1}(\omega x^4_n - 2 x^3_n - \omega x^2_n - 2 x_n)]^{-1}.  \tag{50}
\end{align}

Finally, it is also of interest to note that one can prove the existence of localized excitations, using implicit function theorem, of tilted magnetization or discrete breathers (so called nonlinear localized modes) in a Heisenberg spin chain with easy-plane anisotropy (Zolotaryuk $et~al$. 2001).  It is obvious that there is much scope for detailed study of the discrete spin system to understand magnetic properties, particularly by including Gilbert damping term and also the spin current, see for example a recent study on the existence of vortices and their switching of polarity on the application of spin current (Sheka $et~al$. 2007).
\section{Continuum Spin Systems in (1+1) Dimensions}
\label{}
The continuum case of the LLG equation is a fascinating nonlinear dynamical system.  It has close connection with several integrable soliton systems in the absence of damping in (1+1) dimensions and possesses interesting geometric connections.  Then damping can be treated as a perturbation.  In the (2+1) dimensional case novel structures like line soliton, instanton, dromion, spatiotemporal patterns, vortices, etc. can arise.  They have both interesting mathematical and physical significance.  We will briefly review some of these features and indicate a few of the challenging tasks needing attention.
\subsection{Isotropic Heisenberg Spin System in (1+1) Dimensions}
Considering the (1+1) dimensional Heisenberg ferromagnetic spin system with nearest neighbour interaction, the spin evolution equation without damping can be written as (after suitable scaling)
\begin{align}
\vec S_t = \vec S \times \vec S_{xx},~~~ \vec S = (S^x, S^y, S^z),~~~ \vec S^2 = 1.  \tag{51}
\end{align}
In Eq.(51) and in the following suffix stands for differentiation with respect to that variable.

We now map the spin system (Lakshmanan $et~al$. 1976) onto a space curve (in spin space) defined by the Serret-Frenet equations,
\begin{align}
\vec e_{ix} = \vec D \times \vec e_i, ~~~\vec D = \tau \vec e_1 + \kappa \vec e_3,~~~ \vec e_i.\vec e_i = 1,~~~ i=1,2,3,  \tag{52}
\end{align}
where the triad of orthonormal unit vectors $\vec e_1,\vec e_2,\vec e_3 $ are the tangent, normal and binormal vectors, respectively, and $x$ is the arclength.
Here $\kappa$ and $\tau$ are the curvature and torsion of the curve, respectively, so that $\kappa^2= \vec e_{1x}.  \vec e_{1x} $ (energy density), $\kappa^2 \tau = \vec e_1. (\vec e_{1x} \times \vec e_{1xx}) $ is the current density.

Identifying the spin vector $\vec S(x,t)$ of Eq.(51) with the unit tangent vector $\vec e_1$, from (51) and (52), one can write down the evolution of the trihedral as
\begin{align}
\vec e_{it} = \vec \Omega \times \vec e_i,~~~\vec \Omega = (\omega_1, \omega_2, \omega_3) = \left(\frac {\kappa_{xx}}{\kappa} - \tau^2, -\kappa_x, -\kappa\tau \right).  \tag{53}
\end{align}
Then the compatibility $(\vec e_i)_{xt} = (\vec e_i)_{tx},$   $i$ = 1,2,3, leads to the evolution equation
\begin{align}
\kappa_t &= -2 \kappa_x \tau - \kappa \tau_x ,\tag{54a} \\
\tau_t &= (\frac{\kappa_{xx}}{\kappa}- \tau^2)_x + \kappa \kappa_x , \tag{54b}
\end{align}
which can be rewritten equivalently (Lakshmanan, 1977) as the ubiquitous nonlinear Schr\"odinger (NLS) equation,
\begin{align}
i q_t + q_{xx} + 2 |q|^2 q = 0 ,\tag{55}
\end{align}
through the complex transformation 
\begin{align}
q= \frac{1}{2} \kappa~ exp\left[i \int_{x }^{+\infty} \tau ~d x^{'} \right],  \tag{56}
\end{align}
and thereby proving the complete integrability of Eq.(51).

Zakharow and Takhtajan (1979) have also shown that this equivalence between the (1+1) dimensional isotropic spin chain and the NLS equation is a gauge equivalence.  To realize this, one can write down the Lax representation of the isotropic system (51) as (Takhtajan, 1977) 
\begin{align}
\phi_{1x} = U_1 (x,t,\lambda) \phi_1,~~~ \phi_{1t} = V_1 ( x,t,\lambda) \phi_1,  \tag{57}
\end{align}
where the $(2\times 2)$ matrices $U_1 = i\lambda S,~~~ V_1 = \lambda S S_x + 2i \lambda^2 S,~~~S = \begin{pmatrix} S^z & S^{-}\\S^{+} & -S^z  \end{pmatrix},$
$S^{\pm} = S^x \pm iS^y.$
Then considering the Lax representation of the NLS equation (55), 
\begin{align}
\phi_{2x} = U_2 \phi_2,~~ \phi_{2t} = V_2 \phi_2, \tag{58}
\end{align}
where $U_2 = (A_0 + \lambda A_1),~~~ V_2 = (B_0 + \lambda B_1 + \lambda^2 B_2),$

$A_0 = \begin{pmatrix} 0 & q^{*} \\ -q & 0\end{pmatrix}, ~~~ A_1 = i \sigma_3,~~~ B_0 = \frac{1}{i} \begin{pmatrix} |q^2| & q^*_x \\ q_x & -|q|^2 \end{pmatrix},$

$B_1 = 2A_0,~~~B_2 = 2A_1,~~~\sigma_i $'s are  the Pauli matrices, one can show that with the gauge transformation
\begin{align}
\phi_1 = g^{-1} \phi_2,~~~S= g^{-1} \sigma_3 g,  \tag{59}
\end{align}
(57) follows from (58) and so the systems (51) and (55) are gauge equivalent.

The one soliton solution of the $S^z$ component can be written down as
\begin{align}
S^z(x,t) = 1 - \frac{2\xi }{\xi^2+ \eta^2} ~sech^2 \xi(x-2\eta t-x^0),~~~\xi,\eta,x^0 : constants \tag{60}
\end{align}
Similarly the other components $S^x$ and $S^y$ can be written down and the N-soliton solution deduced.
\subsection{Isotropic chain with Gilbert damping}
The LLG equation for the isotropic case is
\begin{align}
\vec S_t = \vec S \times \vec S_{xx} + \lambda [\vec S_{xx} - (\vec S . \vec S_{xx})\vec S]. \tag{61}
\end{align}
The unit spin vector $\vec S(x,t)$ can be again mapped onto the unit tangent vector $\vec e_1$ of the space curve and proceeding as before (Daniel \& Lakshmanan, 1983) one can obtain the equivalent damped nonlinear $Schr\ddot{o}dinger$ equation,
\begin{align}
iq_t + q_{xx} + 2|q|^2 q = i\lambda [q_{xx} - 2q \int_{-\infty}^{x}(q q^*_{x^{'}} - q^* q_{x^{'}})dx^{'}], \tag{62}
\end{align}
where again $q$ is defined by Eq.(55) with curvature and torsion defined as before. Treating the damping terms proportional to $\lambda$ as a perturbation, one can analyse the effect of damping on the soliton structure.
\subsection{Inhomogeneous Heisenberg Spin System}
Considering the inhomogeneous spin system, corresponding to spatially dependent exchange interaction, 
\begin{align}
\vec S_t = (\gamma_2 +\mu_2 x) \vec S \times \vec S_{xx} + \mu_2 \vec S \times \vec S_x - (\gamma_1 + \mu_1 x)\vec S_x, \tag{63}
\end{align}
where $\gamma_1, ~\gamma_2,~\mu_1$ and $\mu_2$ are constants, again using the space curve formalism, the present author and Robin Bullough (Lakshmanan \& Robin Bullough, 1980) showed the geometrical/gauge equivalence of Eq.(63) with the linearly $x$-dependent nonlocal NLS equation,
\begin{align}
i q_t = ~&i\mu_1 q + i(\gamma_1 + \mu_1 x) q_x \nonumber \\
&+(\gamma_2 + \mu_2 x)(q_{xx} + 2|q|^2 q) + 2\mu_2 (q_x + q \int_{-\infty}^{x}|q|^2 dx^{'}) = 0. \tag{64}
\end{align}
It was also shown in (Lakshmanan \& Bullough, 1980) that the both the systems (63) and (64) are completely integrable and the eigenvalues of the associated linear problems are time dependent.
\subsection{n-dimensional Spherically Symmetric (radial) Spin System}
The spherically symmetric $n$ -dimensional Heisenberg spin system (Daniel $et~al$. 1994)
\begin{align}
\vec S_t(r,t) = \vec S \times \left[\vec S_{rr} + \frac {(n-1)}{r} \vec S_r \right],  \tag{65}\\
\vec S_2(r,t) = 1,~~~\vec S = (S^x,S^y,S^z),~~~ r^2 = r^2_1 + r^2_2+ ...+r^2_n,~~~0\leq r < \infty,  \nonumber
\end{align}
can be also mapped onto the space curve and can be shown to be equivalent to the generalized radial nonlinear $Schr\ddot{o}dinger$ equation,
\begin{align}
i q_t+q_{rr}+ \frac{(n-1)}{r} q_r = \left(\frac {n-1}{r^2} - 2|q|^2 - 4(n-1) \int_0^r \frac{|q|^2}{r^{'}} dr^{'} \right) q. \tag{66}
\end{align}
It has been shown that only the cases $n=1$ and $n=2$ are completely integrable soliton systems (Mikhailov \& Yaremchuk, 1982; Porsezian \& Lakshmanan, 1991) with associated Lax pairs.
\subsection{Anisotropic Heisenberg spin systems}
It is not only the isotropic spin system which is integrable, even certain anisotropic cases are integrable.  Particularly the uniaxial anisotropic chain
\begin{align}
\vec S_t = \vec S \times [\vec S_{xx} + 2A ~S^z \vec n_z + \vec H],~~~\vec n_z = (0,0,1)  \tag{67}
\end{align}
is gauge  equivalent to the NLS  equation (Nakamura \& Sasada, 1982) in the case of longitudinal field $\vec H = (0,0,H^z)$ and is completely integrable.  Similarly the bianisotropic system 
\begin{align}
\vec S_t = \vec S \times \vec J \vec S_{xx},~~~\vec J = diag(J_1,J_2,J_3),~~~J_1 \neq J_2 \neq J_3,\tag{68}
\end{align}
possesses a Lax pair and is integrable (Sklyanin, 1979).  On the other hand the anisotropic spin chain in the case of transverse magnetic field, $H = (H^x,0,0)$, is nonintegrable and can exhibit spatiotemporal chaotic structures (Daniel $et~al$. 1992).

Apart from the above spin systems in (1+1) dimensions, there exists a few other interesting cases which are also completely integrable.  For example, the isotropic biquadratic Heisenberg spin system,
\begin{align}
\vec S_t = \vec S \times \left[S_{xx} + \gamma \left\{\vec S_{xxxx} -\frac{5}{2} (\vec S . \vec S_{xx}) S_{xx} - \frac{5}{3}(\vec S. \vec S_{xxx})\vec S_x \right \}\right] \tag{69}
\end{align}
is an integrable soliton system (Porsezian $et~al$. 1992) and is equivalent to a fourth order generalized nonlinear $Schr\ddot{o}dinger$ equation,
\begin{align}
i q_t + q_{xx} + 2|q|^2 +\gamma \left[ q_{xxxx} + 8 |q|^2 q_{xx} + 2 q^2 q^*_{xx} + 4q|q_x|^2 + 6 q^*q^2_x + 6|q|^4 q \right]= 0 . \tag{70}
\end{align}
Similarly the $SO(3)$ invariant deformed Heisenberg spin equation,
\begin{align}
\vec S_t = \vec S \times \vec S_{xx} + \alpha \vec S_x(\vec S_x)^2, \tag{71}
\end{align}
is geometrically and gauge equivalent to a derivative NLS equation (Porsezian $et~al$. 1987),
\begin{align}
i q_t + q_{xx} + \frac{1}{2} |q|^2 q - i \alpha (|q|^2 q)_x = 0 .\tag{72}
\end{align}
There also exist several studies which maps the LLG equation in different limits to sine-Gordon equation (planar system), Korteweg-de Vries (KdV), mKdV and other equations, depending upon the nature of the interactions.  For details see for example (Mikeska \& Steiner, 1991; Daniel \& Kavitha, 2002).
\section{Continuum Spin Systems in Higher Dimensions}
The LLG equation in higher spatial dimensions, though physically most important, is mathematically highly challenging.  Unlike the (1+1) dimensional case, even in the absence of damping, very few exact results are available in (2+1) or (3+1) dimensions. We briefly point out the progress and challenges.
\subsection{Nonintegrability of the Isotropic Heisenberg Spin Systems}
The (2+1) dimensional isotropic spin system 
\begin{align}
\vec S_t = \vec S \times \left(\vec S_{xx} + \vec S_{yy}\right),~~~\vec S = (S^x,S^y,S^z),~~~\vec S^2 = 1, \tag{73}
\end{align}
under stereographic projection, see Eq.(25), becomes (Lakshmanan \& Daniel, 1981)  
\begin{align}
(1+ \omega \omega^*) [i \omega_t + \omega_{xx} + \omega_{yy}] - 2 \omega^* (\omega^2_x + \omega^2_y ) = 0. \tag{74}
\end{align}
It has been shown (Senthilkumar $et~al$. 2006) to be of non-Painleve nature.  The solutions admit logarithmic type singular manifolds and so the system (73) is nonintegrable.  It can admit special types of spin wave solutions, plane solitons, axisymmetric solutions, etc. (Lakshmanan \& Daniel, 1981).  Interestingly, the static case 
\begin{align}
\omega_{xx} + \omega_{yy} = \frac {2 \omega^*}{(1+\omega \omega^*)^2} (\omega^2_x + \omega^2_y)  \tag{75}
\end{align}
admits instanton solutions of the form
\begin{align}
\omega = (x_1 + ix_2)^m,~~~S^z = \frac{1-(x^2_1+ x^2_2)^m}{1+(x^2_1+x^2_2)^m},~~~m=0,1,2,....  \tag{76}
\end{align}
with a finite energy (Belavin \& Polyakov, 1975; Daniel \& Lakshmanan, 1983).  Finally, very little information is available to date on the (3+1) dimensional isotropic spin systems (Guo \&  Ding, 2008)
\subsection{Integrable (2+1) Dimensional Spin Models}
While the LLG equation even in the isotropic case is nonintegrable in higher dimensions, there exists a couple of integrable spin models of generalized LLG equation without damping in (2+1) dimensions. These include the Ishimori equation (Ishimori, 1984) and Myrzakulov equation (Lakshmanan $et~al$. 1998), where interaction with an additional scalar field is included.
\subsubsection{Ishimori equation:}
\begin{align}
\vec S_t &= \vec S \times (\vec S_{xx} + \vec S_{yy}) + u_x \vec S_y + u_y \vec S_x, \tag{77a}\\
u_{xx} - \sigma^2 u_{yy} &= -2 \sigma^2 \vec S. (\vec S_x \times \vec S_y),~~~\sigma^2 = \pm 1. \tag{77b}
\end{align}
Eq.(77) admits a Lax pair and is solvable by inverse scattering transform (d-bar) method (Konopelchenko \& Matkarimov, 1989).  It is geometrically and gauge equivalent to Davey-Stewartson equation and admits exponentially localized dromion solutions, besides the line solitons and algebraically decaying lump soliton solutions.  It is interesting to note that here one can map the spin onto a moving surface instead of a moving curve (Lakshmanan $et~al$. 1998).
\subsubsection{Myszakulov equation I(Lakshmanan $et~al$. 1998)}
The modified spin equation
\begin{align}
\vec S_t = \left( \vec S \times \vec S_y + u \vec S\right)_x,~~~ u_x = -\vec S.(\vec S_x \times \vec S_y)  \tag{78}
\end{align}
can be shown to be geometrically and gauge equivalent to the Calogero-Zakharov-Strachan equation
\begin{align}
i q_t = q_{xy} + Vq, ~~~ V_x = 2(|q|^2)_y  \tag{79}
\end{align}
and is integrable.  It again admits line solitons, dromions and lumps.

However from a physical point of view it will be extremely valuable if exact analytic structures of the LLG equation  in higher spatial dimensions are obtained and the so called wave collapse problem (Sulem \& Sulem, 1999) is investigated in fuller detail.
\subsection{Spin Wave Instabilities and Spatio-temporal Patterns}
As pointed out in the beginning, the nonlinear dynamics underlying the evolution of nanoscale ferromagnets is essentially described by the LLG equation.  Considering a 2D nanoscale ferromagnetic film with uniaxial anisotropy in the presence of perpendicular pumping, the LLG equation can be written in the form (Kosaka $et~al$. 2005)
\begin{align}
\vec S_t = \vec S \times \vec F_{eff} - \lambda \vec S \times \frac {\partial \vec S}{\partial t},  \tag{80a}
\end{align}
where
\begin{align}
\vec F_{eff} &= J \nabla^2 \vec S + \vec B_a + \kappa S_{\|} \vec e_{\|} + \vec H_m, \tag{80b}\\
\vec B_a &= h_{a\bot}(cos \omega t ~\vec i + sin \omega t ~\vec j) + ha_{\|} \vec e_{\|} \tag{80c}
\end{align}
Here $\vec i, \vec j$ are unit orthonormal vectors in the plane transverse to the anisotropy axis in the direction $\vec e_{\|} = (0,0,1), \kappa$ is the anisotropy parameter, $J$ is the exchange parameter, $\vec H_m$ is the demagnetizing field.  Again rewriting in stereographic coordinates, Eq.(80) can be rewritten (Kosaka $et~al$. 2005) as 
\begin{align}
i(1-i\lambda)\omega_t = J\left(\nabla^2 \omega - \frac{2\omega^* (\nabla \omega)^2}{(1+\omega \omega^*)}\right) - \left( h_{a\|} - \nu + i\alpha \nu + \kappa \frac{(1-|\omega|^2)}{(1+|\omega|^2)}\right) \omega \nonumber \\
+ \frac{1}{2} h_{a\bot} (1-\omega^2) - h_{a\|}\omega + \frac{1}{2} \left(H_m e^{-i\nu t} - \nu^2 H^*_m e^{i\nu t}\right)\omega.  \tag{81}
\end{align}

Then four explicit physically important fixed points (equatorial and related ones) of the spin vector in the plane transverse to the anisotropy axis when the pumping frequency $\nu$ coincides with the amplitude of the static parallel field can be identified.  Analyzing the linear stability of these novel fixed points under homogeneous spin wave perturbations, one can obtain a generalized Suhl's instability criterion, giving the condition for exponential growth of P-modes (fixed points) under spin wave perturbations.  One can also study the onset of different spatiotemporal magnetic patterns therefrom.  These results differ qualitatively from conventional ferromagnetic resonance near thermal equilibrium and are amenable to experimental tests.  It is clear that much work remains to be done along these lines.
\section{Conclusions}
In this article while trying to provide a bird's eye view on the rather large world of LLG equation, the main aim was to provide a glimpse of why it is fascinating both from physical as well as mathematical points of view.  It should be clear that the challenges are many and it will be highly rewarding to pursue them.  What is known at present about different aspects of LLG equation is barely minimal, whether it is the single spin case, or the discrete lattice case or the continuum limit cases even in one spatial dimension, while little progress has been made in higher spatial dimensions.  But even in those special cases where exact or approximate solutions are known, the LLG equation exhibits a very rich variety of nonlinear structures: fixed points, spin waves, solitary waves, solitons, dromions, vortices, bifurcations, chaos, instabilities and spatiotemporal patterns, etc.  Applications are many, starting from the standard magnetic properties including hysterisis, resonances, structure factors to applications in nanoferromagnets, magnetic films and spintronics.  But at every stage the understanding is quite imcomplete, whether it is bifurcations and routes to chaos in single spin LLG equation with different interactions or coupled spin dynamics or spin lattices of different types or continuum system in different spatial dimensions, excluding or including damping.  Combined analytical and numerical works can bring out a variety of new information with great potential applications.  Robin Bullough should be extremely pleased to see such advances in the topic.
\section{Acknowledgements:}
I thank Mr. R. Arun for help in the preparation of this article.  The work forms part of a Department of Science and Technology(DST), Government of India, IRHPA project and is also supported by a DST Ramanna Fellowship.

\end{document}